\documentclass[final,3p,times]{elsarticle}

\usepackage{amsmath}

\newcommand\he[1]{#1^{\dagger}}
\newcommand\gr[1]{\mathrm{#1}}
\DeclareMathOperator{\tr}{Tr}
\newcommand\op{\mathbf\Delta} 
\newcommand\eps{\varepsilon}
\DeclareRobustCommand\openone{\leavevmode\hbox{\small1\normalsize\kern-.33em1}}

\journal{Nuclear Physics A}

\begin{document}

\begin{frontmatter}

\title{Symmetry breaking patterns and collective modes of spin-one color superconductors}
\author[rez]{Tom\'a\v{s} Brauner\fnref{ascr}}
\ead{brauner@th.physik.uni-frankfurt.de}
\fntext[ascr]{On leave from Department of Theoretical Physics, Nuclear Physics
Institute ASCR, CZ-25068 \v{R}e\v{z}, Czech Republic}
\author[ustc]{Jin-yi Pang}
\ead{pangjin@mail.ustc.edu.cn}
\author[ustc]{Qun Wang}
\ead{qunwang@ustc.edu.cn}
\address[rez]{Institute for Theoretical Physics, Goethe University, D-60438 Frankfurt am Main,
Germany}
\address[ustc]{Department of Modern Physics, University of Science and Technology of China,
Anhui 230026, People's Republic of China}

\begin{abstract}
Spin-one color superconductor is a viable candidate phase of dense matter in
the interiors of compact stars. Its low-energy excitations will influence the
transport properties of such matter and thus have impact on late-stage
evolution of neutron stars. It also provides a good example of spontaneous
symmetry breaking with rich breaking patterns. In this contribution, we
reanalyze the phase diagram of a spin-one color superconductor and point out
that a part of it is occupied by noninert states, which have been neglected in
literature so far. We classify the collective Nambu--Goldstone modes, which are
essential to the transport phenomena.
\end{abstract}


\end{frontmatter}


\section{Introduction}
\label{Sec:Intro} Nuclear matter is expected to undergo deconfinement at
extremely high densities such as can be found in the interiors of compact
starts. Due to interactions induced by gluon exchange the resulting quark
matter will behave as a color superconductor (see \cite{Alford:2007xm} for a
recent review). At asymptotically high densities the ground state of quark
matter composed of the three light quark flavors will be the color-flavor
locked (CFL) phase. However, for densities relevant to the phenomenology of
neutron stars where the strange quark mass is large enough, pairing of up and
down quarks only (2SC) is expected to take over. Nevertheless, the constraint
of electric charge neutrality then imposes stress on the pairing by inducing
mismatch between the Fermi surfaces of different flavors. If the mismatch is
too large, cross-flavor pairing becomes impossible, and pairing of quarks with
the same flavor (spin-one pairing) is a candidate phase. Even in the 2SC phase,
the strange quarks left over will still have the possibility to pair with
themselves, resulting again in a spin-one phase.

The spin-one pairing affects the low-energy spectrum of the system since it
gives a gap, albeit small, to quarks that would otherwise remain ungapped. This
gives exponential suppression of transport processes at low temperature. The
knowledge of the phase structure of spin-one color superconductors is therefore
essential for late-stage evolution of neutron stars \cite{Schmitt:2005wg}.
Weak-coupling first-principle calculations \cite{Schmitt:2002sc,Schmitt:2003xq}
show that the single-flavor spin-one color superconductor is in the color-spin
locked (CSL) state. However, when the spin-one pairing of a single flavor is
considered within three-flavor quark matter, for example as a complement to the
primary 2SC pairing, the constraint of color neutrality may lead to other
patterns of spin-one pairing \cite{Alford:2005yy}. We therefore carry out a
phenomenological analysis using the Ginzburg--Landau (GL) theory so as to keep
the full spectrum of possible phases.

\section{Ground state and phase diagram}
The spin-one pair transforms as a color antitriplet and spin triplet. Hence the
order parameter may be written as a $3\times3$ complex matrix $\op$. The
transformation rule then reads $\op\to U\op R^T$, where $U\in\gr{SU(3)_c\times
U(1)_B\equiv U(3)_L}$ and $R\in\gr{SO(3)_R}$. By exploiting this symmetry, the
order parameter can always be cast in the form
\begin{equation}
\op=\begin{pmatrix}
\Delta_1 & ia_3 & -ia_2 \\
-ia_3 & \Delta_2 & ia_1 \\
ia_2 & -ia_1 & \Delta_3
\end{pmatrix},
\label{op_special}
\end{equation}
with real parameters $\Delta_i,a_i$. A detailed analysis shows that there are
eight possible inequivalent phases, differing in the way the continuous
symmetry is spontaneously broken. In addition there are two states,
distinguished only by a discrete symmetry, which occupy a part of the phase
diagram. For the full list of phases the reader is referred to
\cite{Brauner:2008ma}, see also \cite{Schmitt:2004et}. We note that the
symmetry properties of spin-one color superconductors are similar to those of
superfluid Helium 3. Our investigation is to some extent inspired, in both
methods and terminology, by this well understood system
\cite{Vollhardt:1990vw}.

Up to fourth order in $\op$ and two derivatives, the most general
$\gr{U(3)_L\times SO(3)_R}$ invariant GL free energy density functional has the
form
\begin{multline}
\mathcal F[\op]=a_1\tr(\partial_i\op\partial_i\he\op)
+a_2(\partial_i\op_{ai})(\partial_j\op_{aj}^*)
+b\tr(\op\he\op)
+c\eps_{ijk}\op_{ai}^*\partial_j\op_{ak}+\\
+d_1[\tr(\op\he\op)]^2
+d_2\tr(\op\he\op\op\he\op)
+d_3\tr[\op\op^T\he{(\op\op^T)}].
\label{free_energy}
\end{multline}
The sign of the ``mass'' term $b$ determines whether the order parameter is
zero or nonzero. We will hereafter assume that $b<0$ so that we are in the
ordered phase. Note that the $c$ term breaks parity. In ferromagnets it is
responsible for nonuniform, helical ordering in the ground state, as was first
explained by Dzyaloshinsky \cite{Dzyaloshinsky:1958dz} and Moriya
\cite{Moriya:1960mo} fifty years ago. We will therefore refer to it as the DM
term.

\subsection{Phase diagram without parity violation}
\begin{figure}
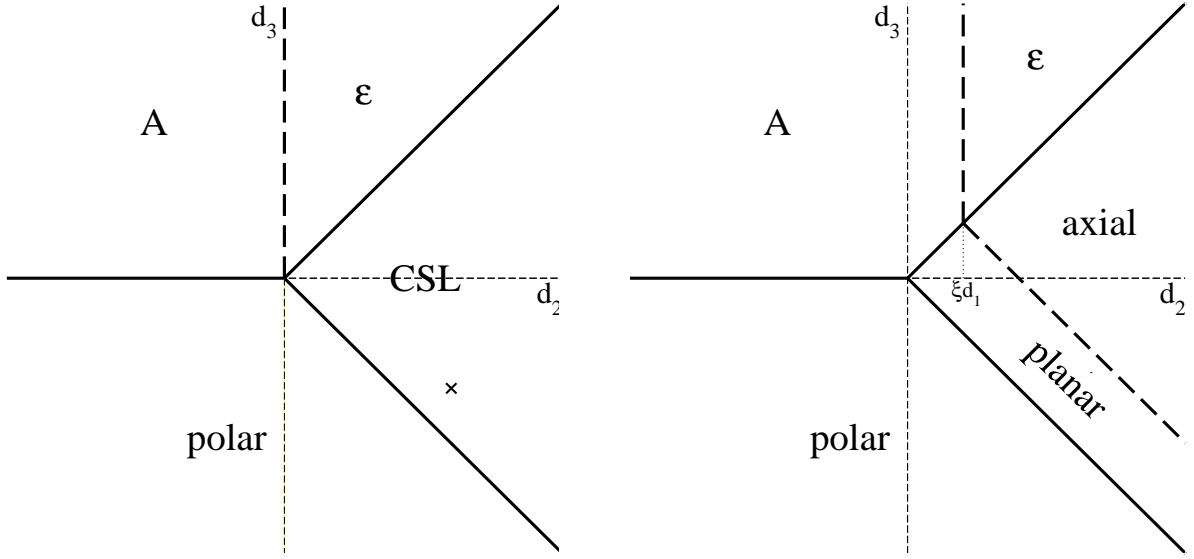

\begin{center}
\includegraphics[width=0.45\textwidth]{phase_diagram}
\hskip2em
\includegraphics[width=0.45\textwidth]{phase_diagram_helix}
\end{center}
\caption{Phase diagram of a spin-one color superconductor in the plane of the
GL parameters $d_2,d_3$. Thick solid and thick dashed lines denote first and
second order phase transitions, respectively. Left panel: without the
parity-violating DM term. The cross indicates the weak-coupling prediction
\cite{Brauner:2008ma}. Right panel: with the parity-violating DM term. The
parameter $\xi$ is defined as $\xi=-c^2/(4a_1b)$.}
\label{Fig:PD}
\end{figure}

Let us first neglect the DM term. The free energy can be minimized by a uniform
order parameter. While its size depends on the balance between the $b$ and
$d_{1,2,3}$ parameters, its orientation in color and spin space is governed
solely by the ratio of $d_2$ and $d_3$. The phase diagram in the $(d_2,d_3)$
plane is plotted in the left panel of Fig.~\ref{Fig:PD}. The four phases that
appear in the phase diagram are defined by the following characteristic shapes
of the order parameter,
\begin{equation}
\op_{\text{CSL}}=\begin{pmatrix}
1 & 0 & 0\\
0 & 1 & 0\\
0 & 0 & 1
\end{pmatrix},\quad
\op_{\text{polar}}=\begin{pmatrix}
0 & 0 & 0\\
0 & 0 & 0\\
0 & 0 & 1
\end{pmatrix},\quad
\op_{\text{A}}=\begin{pmatrix}
0 & 0 & 0\\
0 & 0 & 0\\
1 & i & 0
\end{pmatrix},\quad
\op_\eps=\begin{pmatrix}
0 & 0 & 0\\
0 & 0 & \beta\\
\alpha & i\alpha & 0
\end{pmatrix},
\label{ops_phases}
\end{equation}
the latter two being obtained from the general form \eqref{op_special} by a
left unitary transformation. Note that the same GL analysis for the spin-one
color superconductor was performed more than twenty years ago
\cite{Bailin:1983bm}. However, the authors just compared the energies of the
four inert states (that is, CSL, polar, planar, and A). A proper minimization
of the free energy \eqref{free_energy} reveals that a part of the phase diagram
is occupied by the noninert $\eps$-state.

The finding of a noninert state is one of our main messages. Previous
literature on spin-one color superconductors has rather focused exclusively on
the inert states. It was argued \cite{Alford:2005yy} that while the ground
state of isolated one-flavor quark matter is most likely CSL, color imbalance
induced by the presence of other quark flavors (such as in the ``2SC+$s$''
scheme) may favor the polar phase. Our result opens up the possibility for the
ground state to belong to a whole continuous family of states interpolating
between the CSL and polar ones, the noninert axial phase, which will be
discussed in some detail below. This will lead to a further energy gain and
thus a slight expansion of the region that the spin-one color superconductor
occupies in the phase diagram of dense neutral quark matter.

\subsection{Phase diagram with parity violation}
When the parity-violating DM term is included, the ground state changes
qualitatively. This may be understood by noting that the DM term contains a
single derivative, and leads to a spatially varying ground state. A detailed
analysis \cite{Brauner:2008ma} reveals that the absolute minimum of the free
energy \eqref{free_energy} is either a complex single plane wave, or a real
double (standing) one, depending on the specific phase. Its wavelength is equal
to $4\pi a_1/c$. All plane waves are transverse so that the $a_2$ term does not
play any role. The modified phase diagram is shown in the right panel of
Fig.~\ref{Fig:PD}. The shapes of the five states appearing in the phase diagram
are now, for plane waves propagating in the $z$ direction, given by
\begin{gather}
\notag
\op^{\text{DM}}_{\text{axial}}=\begin{pmatrix}
\alpha\cos kz & \alpha\sin kz & 0\\
-\alpha\sin kz & \alpha\cos kz & 0\\
0 & 0 & \beta
\end{pmatrix},\quad
\op^{\text{DM}}_{\text{planar}}=\begin{pmatrix}
\cos kz & \sin kz & 0\\
-\sin kz & \cos kz & 0\\
0 & 0 & 0
\end{pmatrix},\quad
\op^{\text{DM}}_{\text{polar}}=\begin{pmatrix}
0 & 0 & 0\\
0 & 0 & 0\\
\cos kz & \sin kz & 0
\end{pmatrix},\\
\op^{\text{DM}}_{\text{A}}=e^{ikz}
\begin{pmatrix}
0 & 0 & 0\\
0 & 0 & 0\\
1 & -i & 0
\end{pmatrix},\quad
\op^{\text{DM}}_\eps=\begin{pmatrix}
0 & 0 & 0\\
0 & 0 & \beta\\
\alpha e^{ikz} & -i\alpha e^{ikz} & 0
\end{pmatrix}.
\end{gather}
Note that the change in the topology of the phase diagram: a narrow window of
the planar phase appears, and the CSL phase is replaced by the axial one. The
latter is easily understood by observing that the axial state reduces to CSL
for $\alpha=\beta$. Since only two colors participate in the transverse
nonuniform structure while the third color is uniform and aligned with the $z$
axis, it is natural that a small color imbalance appears. In fact, the two
nonuniformly distributed colors gain energy from the DM term and hence are
favored. As a consequence, in presence of parity violation the ground state of
a spin-one color superconductor is never truly color neutral.

The second and necessary part of our argument is that in dense quark matter
parity violation actually does appear. Quantum chromodynamics itself is parity
conserving, but the DM term can be induced by weak interactions. Another
example of parity violation in a spin-one color superconductor can be found in
the anisotropy of neutrino emission \cite{Schmitt:2005ee}. Since the
electroweak scale is far above the strong one, we expect the DM parameter $c$
to be tiny. Consequently, the ground state of a spin-one color superconductor
will exhibit nonuniform ordering with a very long length scale. A concrete
order-of-magnitude estimate for a typical spin-one color superconductor
\cite{Brauner:2008ma} yields the wavelength of about a millimeter. Of course,
once the $c$ coefficient is small, the energy gain from the formation of the
nonuniform ordering will be tiny. The helical structure will be destroyed by
thermal phonon fluctuations at temperatures higher than about
$0.1\,\text{keV}$. We therefore conclude that the DM ordering will only play
some role in the very late stage of the neutron star life. On the other hand it
should be stressed that the phenomenon is general, the only prerequisites are a
vector order parameter and parity violation.

\section{Collective modes}
Spontaneous symmetry breaking by the nonzero vacuum expectation value of the
order parameter gives rise to collective Nambu--Goldstone (NG) modes. Being
gapless, they dominate low-energy physics of the system, in particular
transport phenomena. (It is nevertheless important to keep in mind that in most
of the spin-one phases, some of the quarks also remain ungapped, or at least
their gap function has nodes on the Fermi surface.) In the following, we will
classify the NG modes in the phases that appear in the phase diagram. We will
for simplicity neglect the DM term which gives rise to parity violation. The
formation of the nonuniform ground state in presence of parity violation brings
additional anisotropy in the low-energy spectrum, leading to non-Fermi liquid
behavior of the system \cite{Kirkpatrick:2005ki}.

Spontaneous symmetry breaking in many-body systems can be intricate provided
some of the conserved charges develop nonzero density
\cite{Nielsen:1975hm,Brauner:2007uw}. The conventional one-to-one
correspondence of NG bosons and broken generators in general does not hold. One
recognizes two qualitatively different classes of NG modes: type-I whose energy
is proportional to an odd (typically first) power of momentum in the
low-momentum limit, and type-II whose energy is proportional to an even
(typically second) power of momentum. One may use the following rule of thumb:
if the commutator of two broken generators develops nonzero vacuum expectation
value, then they give rise to one type-II NG boson, instead of two usual
(type-I) NG bosons. The same phenomenon already appears in the spin-zero 2SC
phase in the Nambu--Jona-Lasinio (NJL) type model with a global color symmetry,
where one of the color charges acquires nonzero density, giving rise to two
type-II NG modes. However, this is just an artifact of the NJL model. Color
neutrality is automatically satisfied in the full gauge theory of strong
interactions and such modes are absent. On the contrary, we will see that some
spin-one color superconducting phases possess physical type-II NG modes coming
from the spontaneously broken rotational symmetry.

Below we list the symmetry-breaking patterns, the unbroken and broken
generators (conveniently defined to be orthogonal), and classify the NG modes
into multiplets of unbroken symmetry. In the general transformation, $\op\to
U\op R^T\equiv(U\otimes R)\op$, we parameterize the left unitary and right
orthogonal matrices as $U(\pi_\alpha)=\exp(i\pi_\alpha\lambda_\alpha)$ and
$R(v_j)=\exp(iv_jM_j)$. Here $\lambda_\alpha$ are the Gell-Mann matrices
extended by $\lambda_0=\sqrt{2/3}\,\openone$, and $(M_j)_{kl}=-i\eps_{jkl}$.
The specific expressions for the (un)broken generators correspond to the forms
of the order parameters in Eq.~\eqref{ops_phases}.

\subsection{CSL phase}
The symmetry breaking pattern is $\gr{U(3)_L\times SO(3)_R\to SO(3)_V}$. The
unbroken subgroup $\gr{SO(3)_V}$ is generated by the diagonal generators,
$(M_j\otimes\openone+\openone\otimes M_j)/\sqrt2$. The nine broken generators
fall into multiplets as follows:
\begin{itemize}
\item $(M_j\otimes\openone-\openone\otimes M_j)/\sqrt2$, $j=1,2,3$. Correspond
to a triplet of type-I NG modes.
\item $\lambda_\alpha\otimes\openone$, $\alpha=1,3,4,6,8$. Correspond to a
$5$-plet of type-I NG modes.
\item $\lambda_0\otimes\openone$. Corresponds a type-I NG singlet.
\end{itemize}
In this phase all charge densities but the $\gr{U(1)_L}$ are zero,
and therefore each broken generator gives rise to one type-I NG
boson. After gauging the color group, eight of the NG bosons are
absorbed by gluons and only the singlet survives in the spectrum,
similar to the CFL phase.

\subsection{Polar phase}
The symmetry-breaking pattern is $\gr{U(3)_L\times SO(3)_R\to
U(2)_L\times SO(2)_R}$. The unbroken symmetry is not locked so that
the left unitary and right rotational groups are broken separately.
The unbroken generators are: $\lambda_{1,2,3}\otimes\openone$
$[\gr{SU(2)_L}]$, $\mathcal P_{12}\otimes\openone$ $[\gr{U(1)_L}]$,
and $\openone\otimes M_3$ $[\gr{SO(2)_R}]$. We have denoted
$\mathcal P_{12}=(\sqrt2\lambda_0+\lambda_8)/\sqrt3$ as the
projector on the first two colors. The generator
$\lambda_8\otimes\openone$ acquires nonzero density, giving rise to
type-II NG bosons in the $\lambda_{4,5,6,7}$ sector. The complete
list of broken generators and the associated NG modes are:
\begin{itemize}
\item $\lambda_\alpha\otimes\openone$, $\alpha=4,5,6,7$. Correspond to an
$\gr{SU(2)_L}$ \emph{doublet of type-II NG bosons}.
\item $\openone\otimes M_j$, $j=1,2$. Correspond to an $\gr{SO(2)_R}$ doublet
of type-I NG modes.
\item $\sqrt2\mathcal P_3\otimes\openone$. Corresponds to a type-I NG singlet.
\end{itemize}
Here $\mathcal P_3=(\lambda_0-\sqrt2\lambda_8)/\sqrt6$ is the projector on the
third color. Altogether there are \emph{7 broken generators}, but only \emph{5
NG bosons}. After gauging the color symmetry, only the $\gr{SO(2)_R}$ vector of
type-I NG bosons survives. They represent two linearly polarized spin waves.

\subsection{A phase}
The symmetry-breaking pattern is $\gr{U(3)_L\times SO(3)_R\to U(2)_L\times
SO(2)_V}$. The unbroken $\gr{U(2)_L}$ is the same as in the polar phase, given
by the $2\times2$ unitary matrices in the upper left corner of $U$. The
unbroken $\gr{SO(2)_V}$ is a diagonal one, generated by
$\sqrt{\frac23}(\mathcal P_3\otimes\openone-\openone\otimes M_3)$. Now the
generators $\lambda_8\otimes\openone$ and $\openone\otimes M_3$ acquire nonzero
density so that there is both color and spin density in the ground state. As a
consequence there are type-II NG bosons in both the color and the spin sector:
\begin{itemize}
\item $\lambda_\alpha\otimes\openone$, $\alpha=4,5,6,7$. Correspond to an
$\gr{SU(2)_L}$ \emph{doublet of type-II NG bosons}.
\item $\openone\otimes(M_1\pm iM_2)/\sqrt2$. Correspond to \emph{one type-II NG
boson} of $\gr{SO(2)_V}$.
\item $\sqrt{\frac23}(\mathcal P_3\otimes\openone+\openone\otimes M_3)$.
Corresponds to a type-I NG singlet.
\end{itemize}
There are \emph{7 broken generators}, but only \emph{4 NG bosons}.
Only the type-II NG singlet survives after gauging the color group.
It represents a circularly polarized spin wave.

\subsection{$\eps$ phase}
The symmetry-breaking pattern is $\gr{U(3)_L\times SO(3)_R\to U(1)_L\times
SO(2)_V}$, with the unbroken generators given by $\sqrt2\mathcal
P_1\otimes\openone$ $[\gr{U(1)_L}]$, and $\sqrt{\frac23}(\mathcal
P_3\otimes\openone-\openone\otimes M_3)$ $[\gr{SO(2)_V}]$. Here $\mathcal P_1$
is obviously the projector on the first color. The generators
$\lambda_{3,8}\otimes\openone$ and $\openone\otimes M_3$ acquire nonzero
density. Thanks to low unbroken symmetry the collective excitation spectrum has
a rich structure, in particular there are no nontrivial multiplets of states:
\begin{itemize}
\item $\openone\otimes\frac1{\sqrt2}(M_1\pm iM_2)$. Correspond to \emph{one type-II
NG boson} of $\gr{SO(2)_V}$.
\item $\frac1{\sqrt2}(\lambda_{1,4,6}\pm i\lambda_{2,5,7})\otimes\openone$. Correspond
to \emph{three type-II NG singlets}.
\item $\sqrt2\mathcal P_2,
\sqrt{\frac23}(\mathcal P_3\otimes\openone+\openone\otimes M_3)$. Correspond to
two type-I NG singlets.
\end{itemize}
Altogether there are \emph{10 broken generators}, but only \emph{6 NG bosons}.
All gluons acquire mass by the Higgs mechanism, and the only NG boson that
remains after gauging the color symmetry is again the circularly polarized spin
wave.

\subsection{Mode mixing at nonzero momentum}
It is interesting to note that in absence of the $a_2$ (and $c$) term, the
free energy \eqref{free_energy} is invariant under separate spacetime rotations
and orthogonal $\gr{SO(3)_R}$ rotations of the order parameter. It is only the
$a_2$ term which ties these two symmetries together and makes the order
parameter a space three-vector. Some of the NG bosons listed above
therefore stem from spontaneous breaking of a spacetime symmetry. This has an
intriguing consequence. Our classification of the NG modes was based on the
symmetry breaking by a uniform order parameter. However, when fluctuations of
the order parameter, that is, excitations above the ground state, are
considered, their nonzero momentum leads to further breaking of the rotation
symmetry. Consequently, some of the modes belonging to different multiplets
as defined above may mix. Strictly speaking our classification is 
only accurate in the long-wavelength limit. A full treatment
of the general case will be reported in a forthcoming publication.

\section{Conclusions}
In analogy to the coupling of orbital angular momentum and spin in atomic
physics, the color--spin coupling is a new kind of interaction involving
subatomic degrees of freedom and appears in the spin-one color
superconductivity. It provides a perfect example of spontaneously symmetry
breaking in a condensed matter system with its rich symmetry breaking patterns.
We analyzed the phase diagram of a spin-one color superconductor and pointed
out that a part of it is occupied by noninert states, which have been neglected
in literature so far. We classified the collective excitations (the NG modes)
in a systematic way and obtained their dispersion relations. The details of our
calculations will be published elsewhere.

\section*{Acknowledgments}
TB is supported in part by the Alexander von Humboldt Foundation, and by the
ExtreMe Matter Institute EMMI in the framework of the Helmholtz Alliance
Program of the Helmholtz Association (HA216/EMMI). QW is supported in part by
the `100 talents' project of Chinese Academy of Sciences (CAS) and by the
National Natural Science Foundation of China (NSFC) under the grants 10675109
and 10735040.

\bibliography{helix}

\begin{thebibliography}{10}
\expandafter\ifx\csname url\endcsname\relax
  \def\url#1{\texttt{#1}}\fi
\expandafter\ifx\csname urlprefix\endcsname\relax\def\urlprefix{URL }\fi
\expandafter\ifx\csname href\endcsname\relax
  \def\href#1#2{#2} \def\path#1{#1}\fi

\bibitem{Alford:2007xm}
M.~G. Alford, A.~Schmitt, K.~Rajagopal, T.~Schafer, {Color superconductivity in
  dense quark matter}, Rev. Mod. Phys. 80 (2008) 1455--1515.
\newblock \href {http://arxiv.org/abs/0709.4635 [hep-ph]}
  {\path{arXiv:0709.4635 [hep-ph]}}.

\bibitem{Schmitt:2005wg}
A.~Schmitt, I.~A. Shovkovy, Q.~Wang, {Neutrino emission and cooling rates of
  spin-one color superconductors}, Phys. Rev. D73 (2006) 034012.
\newblock \href {http://arxiv.org/abs/hep-ph/0510347}
  {\path{arXiv:hep-ph/0510347}}.

\bibitem{Schmitt:2002sc}
A.~Schmitt, Q.~Wang, D.~H. Rischke, {When the transition temperature in color
  superconductors is not like in BCS theory}, Phys. Rev. D66 (2002) 114010.
\newblock \href {http://arxiv.org/abs/nucl-th/0209050}
  {\path{arXiv:nucl-th/0209050}}.

\bibitem{Schmitt:2003xq}
A.~Schmitt, Q.~Wang, D.~H. Rischke, {Electromagnetic Meissner effect in
  spin-one color superconductors}, Phys. Rev. Lett. 91 (2003) 242301.
\newblock \href {http://arxiv.org/abs/nucl-th/0301090}
  {\path{arXiv:nucl-th/0301090}}.

\bibitem{Alford:2005yy}
M.~G. Alford, G.~A. Cowan, {Single-flavour and two-flavour pairing in
  three-flavour quark matter}, J. Phys. G32 (2006) 511--528.
\newblock \href {http://arxiv.org/abs/hep-ph/0512104}
  {\path{arXiv:hep-ph/0512104}}.

\bibitem{Brauner:2008ma}
T.~Brauner, {Helical ordering in the ground state of spin-one color
  superconductors as a consequence of parity violation}, Phys. Rev. D78 (2008)
  125027.
\newblock \href {http://arxiv.org/abs/0810.3481 [hep-ph]}
  {\path{arXiv:0810.3481 [hep-ph]}}.

\bibitem{Schmitt:2004et}
A.~Schmitt, {The ground state in a spin-one color superconductor}, Phys. Rev.
  D71 (2005) 054016.
\newblock \href {http://arxiv.org/abs/nucl-th/0412033}
  {\path{arXiv:nucl-th/0412033}}.

\bibitem{Vollhardt:1990vw}
D.~Vollhardt, P.~W{\"o}lfle, {The Superfluid Phases of Helium 3}, Taylor and
  Francis, 1990.

\bibitem{Dzyaloshinsky:1958dz}
I.~Dzyaloshinsky, {A thermodynamic theory of ``weak'' ferromagnetism of
  antiferromagnetics}, J. Phys. Chem. Solids 4 (1958) 241.

\bibitem{Moriya:1960mo}
T.~Moriya, {Anisotropic superexchange interaction and weak ferromagnetism},
  Phys. Rev. 120 (1960) 91.

\bibitem{Bailin:1983bm}
D.~Bailin, A.~Love, {Superfluidity and Superconductivity in Relativistic
  Fermion Systems}, Phys. Rept. 107 (1984) 325.

\bibitem{Schmitt:2005ee}
A.~Schmitt, I.~A. Shovkovy, Q.~Wang, {Pulsar kicks via spin-1 color
  superconductivity}, Phys. Rev. Lett. 94 (2005) 211101.
\newblock \href {http://arxiv.org/abs/hep-ph/0502166}
  {\path{arXiv:hep-ph/0502166}}.

\bibitem{Kirkpatrick:2005ki}
T.~R. Kirkpatrick, D.~Belitz, {Nonanalytic corrections to Fermi-liquid behavior
  in helimagnets}, Phys. Rev. B72 (2005) 180402.
\newblock \href {http://arxiv.org/abs/cond-mat/0506770}
  {\path{arXiv:cond-mat/0506770}}.

\bibitem{Nielsen:1975hm}
H.~B. Nielsen, S.~Chadha, {On How to Count Goldstone Bosons}, Nucl. Phys. B105
  (1976) 445.

\bibitem{Brauner:2007uw}
T.~Brauner, {Goldstone bosons in presence of charge density}, Phys. Rev. D75
  (2007) 105014.
\newblock \href {http://arxiv.org/abs/hep-ph/0701110}
  {\path{arXiv:hep-ph/0701110}}.

\end{thebibliography}
\bibliographystyle{elsarticle-num}

\end{document}